\documentclass[11pt]{amsart}
\usepackage{amsmath}
\usepackage{amssymb}
\usepackage{graphicx}
\usepackage{caption}
\usepackage{subcaption}
\usepackage{siunitx}

\usepackage[acronym]{glossaries}
\newacronym{tb}{TB} {tight-binding}
\newacronym{npg}{NPG} {nanoporous graphene}
\newacronym{gnr}{GNR} {graphene nanoribbon}
\newacronym{gnrs}{GNRs} {graphene nanoribbons}
\newacronym{od}{1D}{one dimensional}
\newacronym{ls}{LS}{large-scale}
\newacronym{dft}{DFT}{Density Functional Theory}
\newacronym{ef}{$E_F$}{Fermi Energy}
\makeatletter
\g@addto@macro{\endabstract}{\@setabstract}
\newcommand{\authorfootnotes}{\renewcommand\thefootnote{\@fnsymbol\c@footnote}}%
\makeatother

\begin{document}
\begin{center}
  \LARGE 
  \textbf{Field-controlled Electronic Breathing Modes and Transport in Nanoporous Graphene} \par \bigskip

  \normalsize
  \authorfootnotes
  Alan Ernesto Anaya Morales\footnote{alemor@dtu.dk}\textsuperscript{1}, 
  Mads Brandbyge\footnote{mabr@dtu.dk}\textsuperscript{1},
  \bigskip

  \textsuperscript{1}Department of Physics,Technical University of Denmark Fysikvej, 309, 2800 Kgs. Lyngby, Denmark\par

  \today
\end{center}
\begin{abstract}
       Nanoporous graphene (NPG) has been fabricated by on-surface-self assembly in the form of arrays of $\sim 1$ \si{\nano\meter}-wide graphene nanoribbons connected via molecular bridges in a two-dimensional crystal lattice. It is predicted that NPG may, despite its molecular structure, work as electron waveguides that display e.g. Talbot wave interference. Here, we demonstrate how the electronic wave guidance may be controlled by the use of electrical fields transverse to the ribbons; at low fields, point-injected currents display spatially periodic patterns along the ribbons, while high fields localize the injected current to single ribbons. This behavior constitutes an electronic version of optical breathing modes of Bloch oscillations, providing a simple mechanism for controlling the current patterns down to the molecular scale. The robustness of the self-repeating patterns under disorder demonstrate that the breathing modes of single-ribbon injections offer exciting opportunities for applications in nanoelectronics, molecular sensing, and quantum information processing.  
\end{abstract}

\section{Introduction}
On-surface synthesis of \gls{npg} provides a versatile platform for the development of semiconducting graphene devices, offering a wide range of potential applications in nanoelectronics, molecular sensing, and quantum information processing \cite{BottomUpNPG,PhotonicCrystal,sensing,quantum}. The unique topology of \gls{gnrs} in \gls{npg} results in a pronounced in-plane anisotropy, enabling energy-dependent quasi-\gls{od} localization of electron states near the conduction band even in stacked structures \cite{DiazdeCerio25}. These features position \gls{npg} as a potential material for spatially controlling electron transport on the molecular scale.

Despite these features, \gls{ls} transport simulations, based on parameter-free tight binding (TB) Hamiltonians, reveal that inter-ribbon coupling disrupts longitudinal electron confinement, causing the single-point injected current to spread over tenths of nanometers \cite{TalbotNPG}. This behavior manifests itself as Talbot interference, a phenomenon commonly observed in optical gratings, ultra-cold atoms in optical lattices, photonic crystals, optical wave-guides, and plasmonic systems \cite{PhotonicCrystal,Talbotoptics,ultracoldatoms,Talbot_wave_guides,Talbot_Discrete,Talbot_plasmonic}. The current spread reduces \gls{npg}'s effectiveness for high-precision applications where control of localized currents is desirable. To address this challenge, promising strategies, such as quantum interference engineering \cite{QuantumInterference} and hetero-atom doping \cite{hNPG,ElectrochemicalNPG} have been explored. However, approaches based on structural changes present experimental challenges and limit tunability. This highlights the need to explore alternative, non-chemical methods for spatially controlling electron transport in NPG.

In this work, we investigate the effect of applied constant transverse electric fields on electron transport in \gls{npg}.  We found that such fields induce periodic modulation in the spatial distribution of point-injected currents in the form of Bloch oscillations featuring breathing modes \cite{Bloch1929,Zener,OpticalBloch,stronglycorrelated,Zener}. We present a comprehensive investigation by combining \gls{ls} transport simulations  (Fig.\ref{fig:figure1}) with an analytical model based on a discrete differential equation (DDE) akin to optical wave guides with graded refractive index \cite{OpticalBloch}. We find that the resulting field-tunable current patterns are robust, even in the presence of static disorder, revealing a practical route for dynamic control of electron flow in 2D graphene nanostructures.

\section{Computational Methods}
\begin{figure*}[ht!]
    \centering
    \includegraphics[width=1\linewidth]{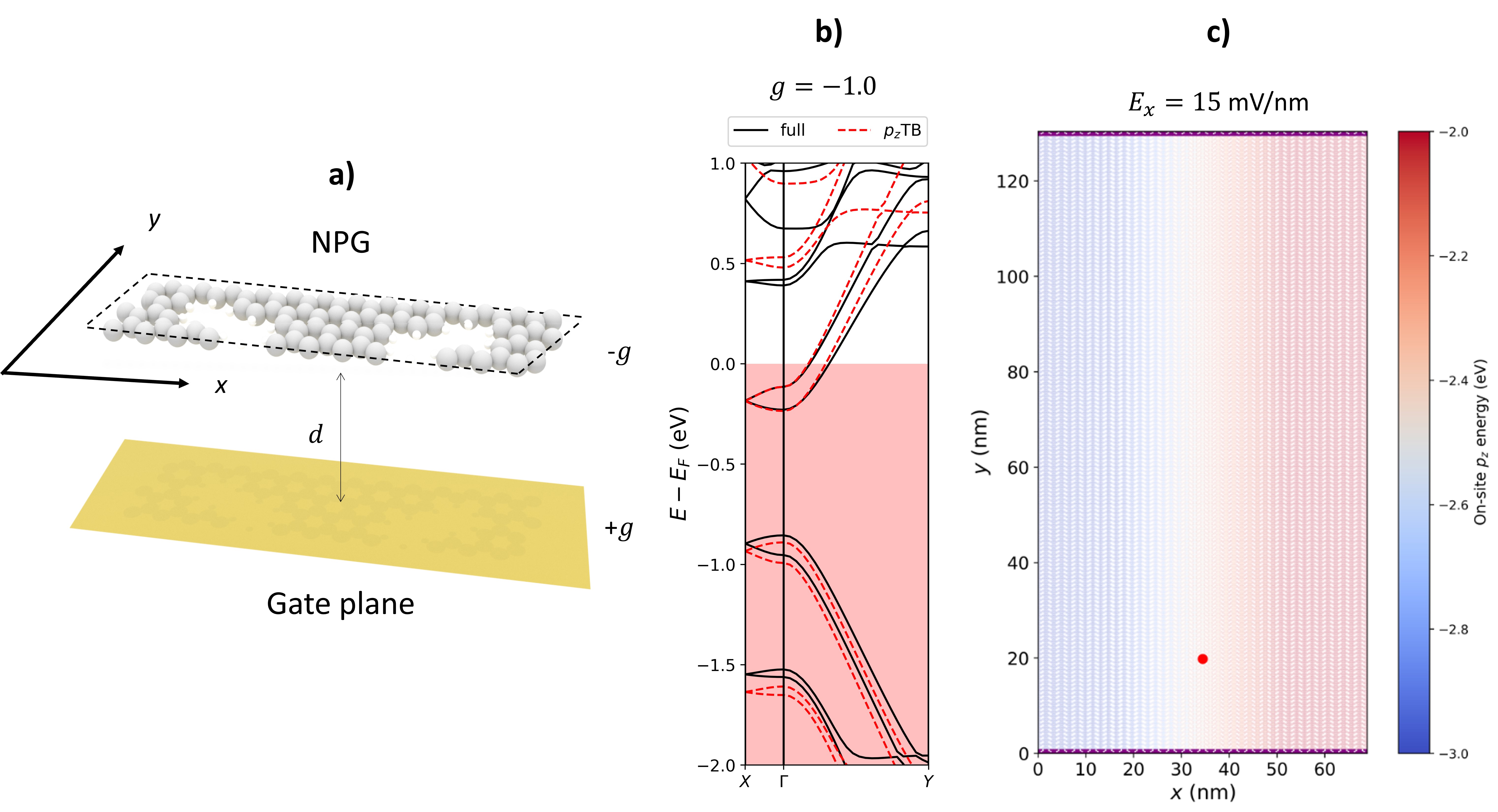}
    \caption{\textbf{From DFT to Large-Scale Transport.} (a) Schematic of a field-effect gate setup on  \gls{npg}. The periodic unit-cell contains two GNRs and the carbon atoms at the pores are passivated with hydrogen. Electrostatic gates are used to increase the carrier density in NPG by $n_g=g\times10^{13}$ $\text{e}/\text{cm}^2$. The gate plane is placed at a distance, $d$, of 15 \si{\nano \meter} below NPG. (b)  Band structure comparison of a $n$-doped NPG from DFT (full black lines) and  $p_z$-only TB (red dashed lines) Hamiltonians. (c) Schematic of a \gls{ls} device composed from $21\times150$ repeated unit cell of NPG with a field in the transverse ribbon ($x$) direction. The field is modeled here as a linear ramp in the $p_z$ on-site energies of the carbon atoms. Bottom and top electrodes are marked in purple, and the injection point is indicated by a red dot.}
    \label{fig:figure1}
\end{figure*}

To study electron transport, we employed an effective \gls{tb} model, which reproduces the full DFT electronic structure relevant for transport, following the method used in previous studies \cite{TalbotNPG, QuantumInterference}, (Fig. \ref{fig:figure1}a-b). Electron transport is calculated using Non-Equilibrium Green's function methods, where point-injection of current at a single atomic site is included as a self-energy in the Green's function to mimic atomic contact to an STM tip electrode \cite{LS}. This approach balances computational cost and accuracy, making it feasible to study systems on the scale of hundreds of nanometers. The \gls{ls} setup is shown in Figure \ref{fig:figure1}c.

\subsection{DFT calculations}
The electronic structure was calculated using \gls{dft}, with the SIESTA package \cite{SIESTA}, employing the Generalized Gradient Approximation (GGA) with the Perdew-Burke-Ernzerhof (PBE) exchange correlation functional \cite{PBE}, a DZP base set, and a Monkhorst-Pack k-point grid of $5\times30\times1$, ensuring convergence of the electronic band structure of \gls{npg}. 

To simulate electrostatic gating, a capacitor model was included in the DFT calculations \cite{Gating}. The electrostatic-gate setup is schematized in Figure \ref{fig:figure1}a. The gate plane was placed at a distance $d$ of $15$~{\AA} below \gls{npg}.  Electrostatic gating in NPG results in a rigid band shift \cite{BottomUpNPG}, which allows doping of the material. Charge densities correspond to $n_g=g\times10^{13}
$ $\text{e}/\rm{cm}^{2}$. For $g<0$ ($g>0$), systems are $n$($p$)-doped. Due to the electron-hole symmetry of \gls{npg}, the same behavior is observed for $n$- and $p$-type doping. Here, we present the results for $n$-doping ($g=-1$,  Fig. \ref{fig:figure1}b). Slab-dipole corrections were included in gated calculations to cancel artificial electric fields that arise due to periodic boundary conditions along $z$. The unit cell included a vacuum of $50$ \AA,  to avoid spurious image interactions. 

\subsection{Tight-Binding Model for Large-Scale Transport}
\begin{figure}[ht!]
\centering
\includegraphics[width=1\linewidth]{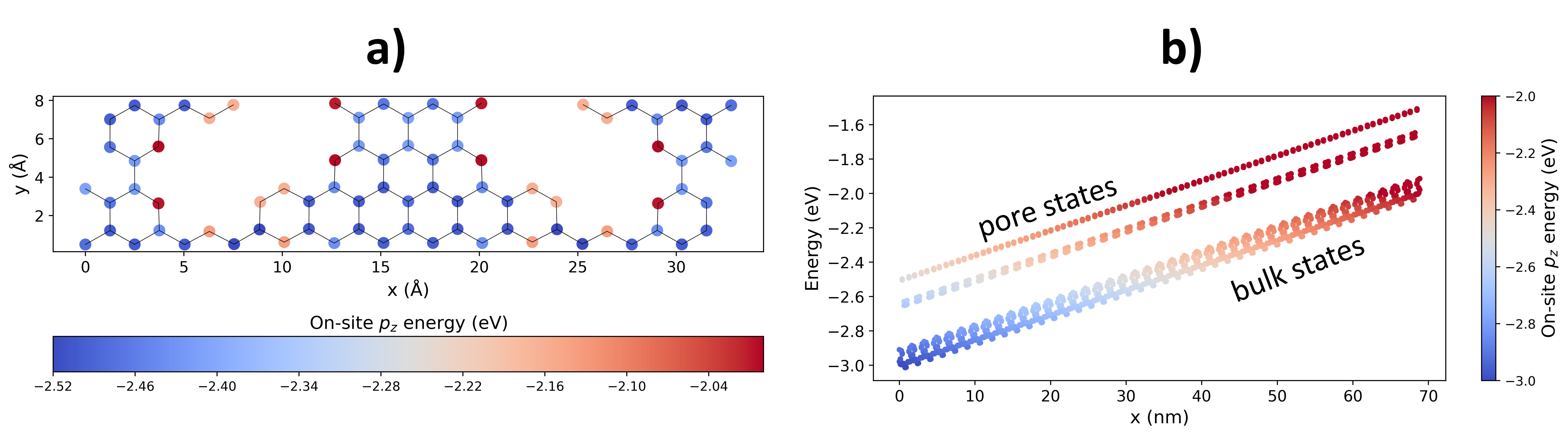}\caption{\textbf{Electric-field induced potential drop model} (a) On-site $p_z$ energies ($\epsilon_{i}$) projected onto the carbon atoms in $n$-doped ($g=1$) \gls{npg} unit-cell obtained from a \gls{dft} pruned  Hamiltonian.  (b) On-site energy ramp along the transverse direction in the \gls{ls} \gls{npg} device for a field of $E_x=15$ \si{\milli \volt}/\si{\nano\meter}. The $p_z$ states on carbon atoms in the bulk GNRs form a \textit{Wannier–Stark} ladder at lower energy compared to the $p_z$-states from the pores and the bridges.}
\label{fig:figure2}
\end{figure}

A nonorthogonal TB-like Hamiltonian is obtained by pruning the \gls{dft} Hamiltonian using the SISL package \cite{sisl}, retaining only the $p_z$ orbitals of the carbon atoms (Fig. \ref{fig:figure1}b). The diagonal elements of the resulting Hamiltonian correspond to effective on-site energies, $\epsilon_i$. These are shown in Figure \ref{fig:figure2}a.  Differences in on-site energies arise from the chemical environments felt by the carbon atoms sitting at the edge, pores, or at the bulk of the \gls{gnrs} in \gls{npg}.

A \gls{ls} device is constructed by tiling the (TB-like) Hamiltonian from a primitive NPG cell. This allows us to reach device dimensions up to hundreds of nanometers. Here we consider devices of \SI{69}{\nano\meter} $\times $ \SI{130}{\nano\meter}, which contain about 25200 atoms, by tiling the unit cell by $21\times150$. We used bottom and top electrodes and complex absorption potential (CAP) regions on the left and right sides of the device (CAP width of 50 \si{\nano\meter}), to absorb electrons without reflection at the boundaries of the device. This provides a description of an infinite surrounding NPG environment \cite{LS, CAP}. 

A central point of the present study is to consider the inclusion of an electrical field transverse to the ribbons, as depicted in Figure \ref{fig:figure1}c. This was inspired by previous work on graphene field-effect transistors (GFET), where a graded-potential gate enabled a linearly varying electric field, and allowed for spatial electronic modulation of graphene \cite{linear,Lassaline_2024}. To model this field, we apply a linear ramp by shifting the on-site energy for an atom $i$ at position $x_i$ according to,
\begin{eqnarray}
   \epsilon_i(x)= \epsilon_{i} +  U\frac{x_i- L_{x}/2}{L_x} \label{eq:onsite}
\end{eqnarray}
where $U$= e $V_x$ represents the total potential drop along the device, and $L_x$ is the device width, as shown in Figure \ref{fig:figure2}(b). The applied field corresponds to $E_x=U/e L_x$. For the given device geometry, the potential drop of $U=1.0$ e\si{ \volt} corresponds to fields of $E_x=15$ \si{\milli \volt}/\si{\nano\meter}.

\begin{figure}[ht!]
\centering
\begin{subfigure}{0.8\textwidth}
\includegraphics[width=\linewidth]{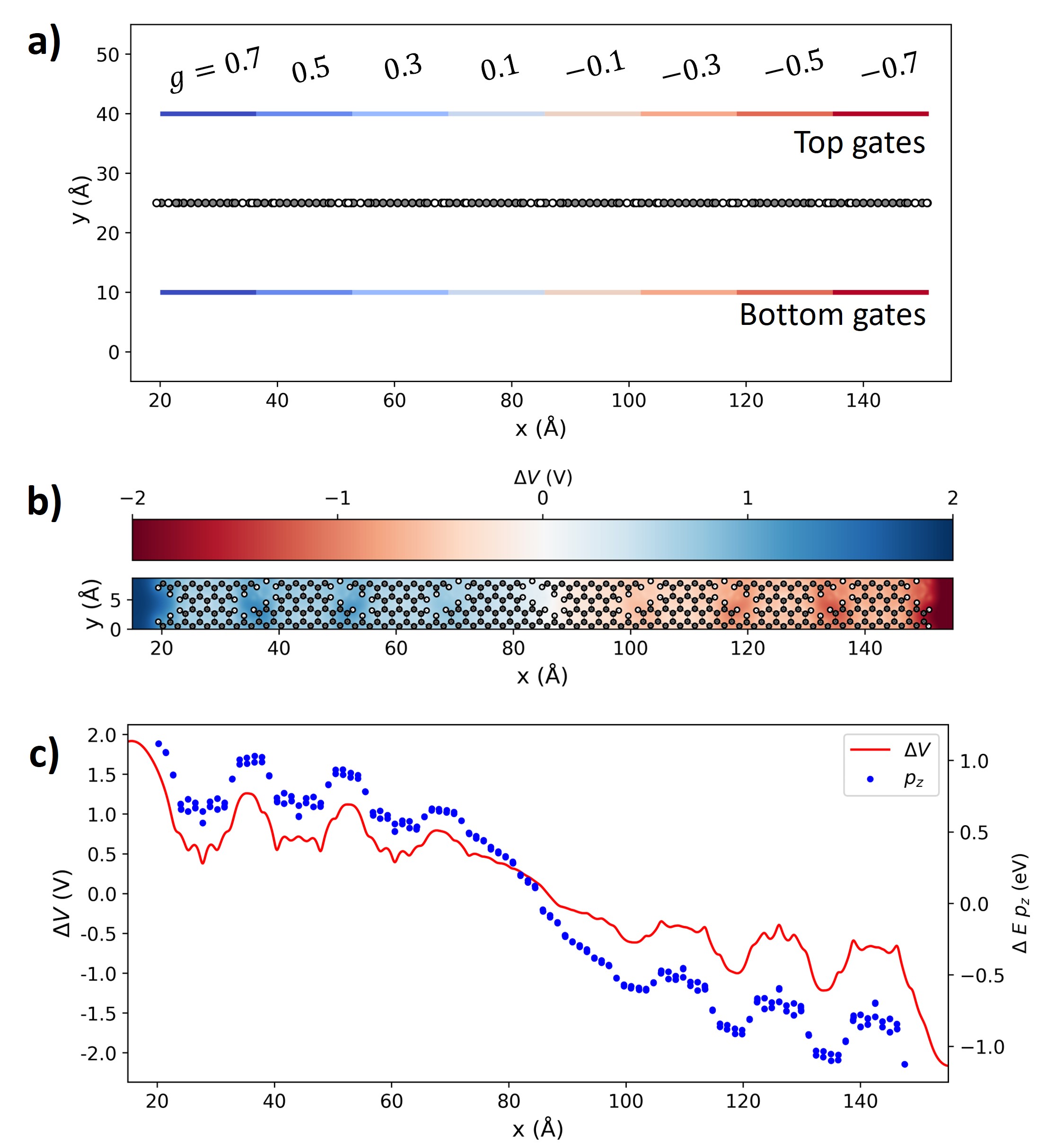}
\end{subfigure}
\caption{\textbf{DFT calculation of graded double-gated NPG.}  a) Discrete double graded-potential gate model where each ribbon sees a different gate ($g$). The top/bottom gates are placed at a distance of $\sim 15$ \AA above/ below the \gls{gnrs}. The charge density  in the gate ($n_g$) is varied linearly in steps from $g=0.7$ to $g=-0.7$ on both top and bottom gates. (b)  Change in electrostatic potential with gating, $\Delta V$, in the $xy$-plane containing the \gls{npg} carbon atoms ($z=25$ \AA). c)  Electrostatic potential ($z=25$ \AA)  along the $x$ direction at the center of the GNRs in the $y$ direction. This correlates to a change in the $p_z$  on-site energies, shown as  blue dots.}
\label{fig:figure3}
\end{figure}

To validate the linear ramp model used in the \gls{ls} calculations, we performed DFT calculations on a much smaller test system that incorporates a ramped gate, as depicted in Fig.~\ref{fig:figure3}a. To avoid problems with periodic field images in the periodic DFT cell, we considered a double gate with top and bottom gates placed at $15$~\si{\angstrom} above/below the NPG plane. However, the results presented here are general for both double and single gate devices. To mimic the ramp itself, the graded-potential gates were discretized into sections of fixed charge density going from $p$ to $n-$doping.  We considered systems of $8$ \gls{gnrs} repeated along the $x$ direction and a vacuum to allow left-right polarization, and with periodic boundary conditions along the longitudinal $y$ direction. The left-right edges were passivated with hydrogen atoms. For computational efficiency, the calculations were performed with a SZP basis set, which captures the features of the low energy bands in \gls{npg} well. 
The calculations showed that the electrostatic potential along the device $\Delta V$, exhibits a modulated drop, as shown in Figure~\ref{fig:figure3}b. The change in the on-site energies of $p_z$ states, exhibit the same behavior. We note that the fields and potential drop are exaggerated in the small DFT model, much due to the small size and edge effects. However, the calculation shows a linear potential drop in the middle of the structure, even for this small system size and strong field, and thus justifies the use of a linear ramp model for the larger system.

\subsection{Transport Simulations}
Transport calculations were performed using the Green's function method as implemented in TBTrans \cite{transiesta}.  The Green's function of the system is computed as:
\begin{eqnarray}
\mathbf{G}=\left [ \mathbf{S}(z)-\mathbf{H}(z) - \sum_i \mathbf{\Sigma}_i(z)\right]^{-1}\label{eq:GF} 
\end{eqnarray}
where $z=E+i\eta$ is the complex energy parameter, $\mathbf{S}$ and $\mathbf{H}$ are the overlap and Hamiltonian. The $\mathbf{\Sigma}_i$ are self-energies that describe the connection of the device to electrodes, the CAP's \cite{CAP} and, especially, the point-injector electrode \cite{LS}. Transport simulations were performed at injection energies $E-E_F=0.2$ e\si{\volt} to investigate the linear dispersion regime of the conduction band (cf. Fig.~\ref{fig:figure1}b). Calculations for different gatings showed that linear dispersion is required to observe breathing modes (Supplementary Material).

The point-injected current is visualized from the bond-transmission (or spectral bond-current) maps. Bond transmissions are given by,
\begin{eqnarray}
  T_{ij}(E)=i[(\mathbf{H}_{ji}-E\mathbf{S}_{ji})\mathbf{A}_{ij}(E)-(\mathbf{H}_{ij}-E\mathbf{S}_{ij})\mathbf{A}_{ji}(E)]] \label{eq:Tij}
\end{eqnarray}
representing the probability distribution of electrons passing through specific bonds \cite{Bondcurrent_Nakanishi2001,Bondcurrent_Solomon2010}. For atoms with more than one orbital we can define the atom transmission which corresponds to a scalar that measures the transmission or spectral current flowing through the atom ($i$) by summing over all atoms ($j$) to which there is electronic coupling  \cite{transiesta},
\begin{eqnarray}
   T_i(E)=\frac{1}{2}\sum_{i,j}T_{ij}(E)\label{eq:Ti}
\end{eqnarray}
For our particular case, in which each carbon atom is represented by single $p_z$ orbital, these terms are equivalent.  Atom transmission maps are presented with color-bars corresponding to values from $T_{\text{min}}=0$ to $T_{\text{max}}=0.001$.
\section{Field-Induced Breathing Modes}

\begin{figure*}[ht!]
\centering
\includegraphics[width=\linewidth]{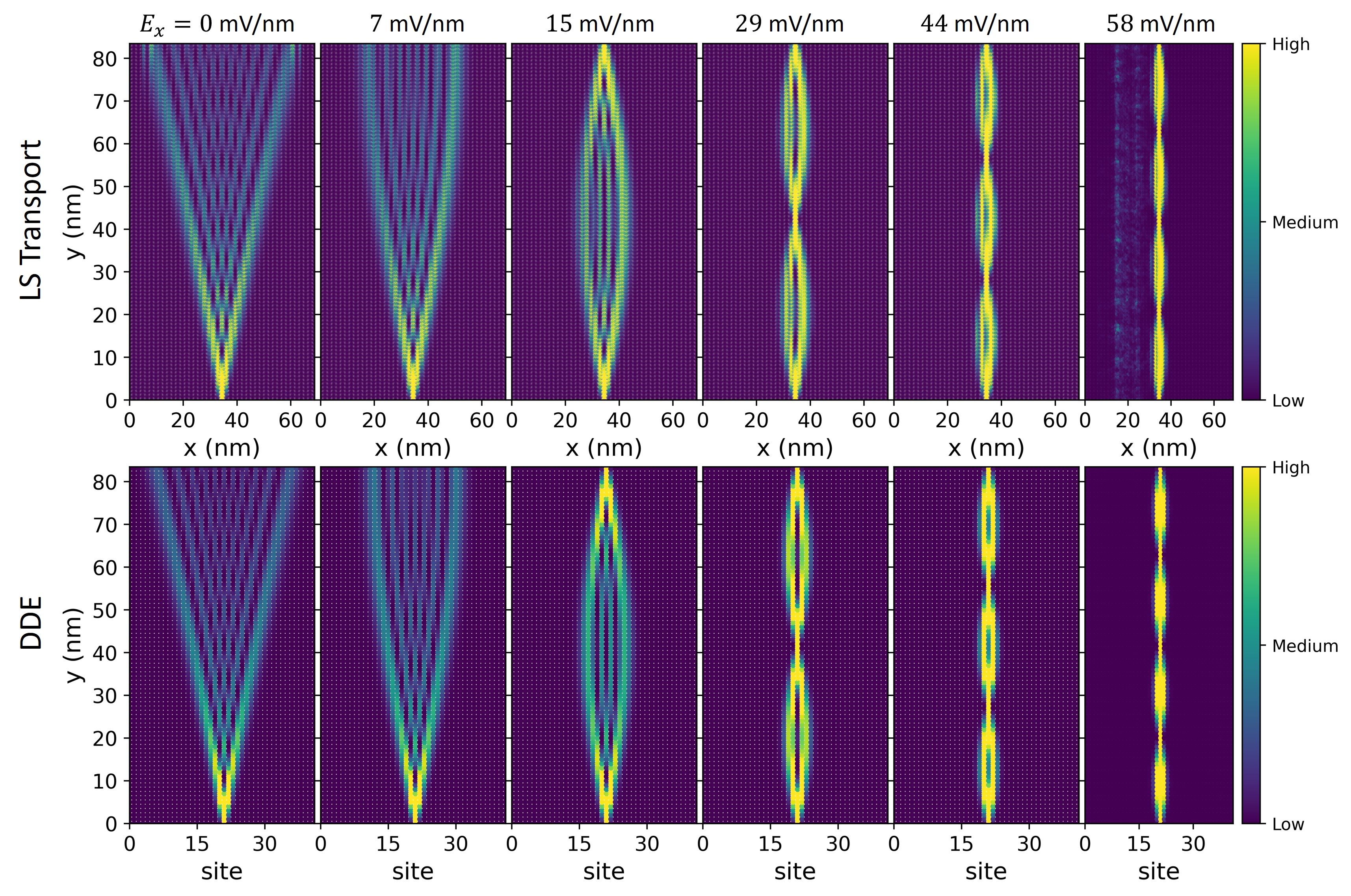}
    \caption{\textbf{Bloch Oscillations in \gls{npg} for a constant transverse field}. Comparison between atom transmission ($T_i$) from \gls{ls} transport calculations of atom transmission (top panels) and  Discrete Differential Equation (DDE, lower panels) for the probability $|\psi_n|^2$ ($n$ ribbon). The field $E_x$ was varied from   $E_x=0.0$ \si{\milli \volt}/\si{\nano\meter} up to $E_x=58$ \si{\milli \volt}/\si{\nano\meter}. $T_i$ are shown from 0 to 0.001. In the DDE, the initial intensity $|\psi(z_0)|^2$ is such to reproduce the observed bond currents. For the \gls{ls}, we observe a faint asymmetry due to Landau-Zener tunnelling into the bridge/pore states (top, rightmost panel).}
    \label{fig:Bloch oscillations}
\end{figure*}

 Figure \ref{fig:Bloch oscillations} (top panel) shows the atom-resolved current maps ($T_i$) for different transverse field strengths calculated with the \gls{ls} method. The transverse field, $E_x$, was varied from $0.0$ \si{\volt}/\si{\nano\meter} to $58$ \si{\milli \volt}/\si{\nano\meter}.
 In the absence of a field, the injected current exhibits Talbot interference \cite{TalbotNPG}, with characteristic spatial diffraction.  For low applied fields ($E_x \approx  7$ \si{\milli \volt}/\si{\nano\meter}) the current begins to refocus downstream $\sim 80$ \si{\nano\meter} from the source. For a field twice as large ($E_x \approx$  15 \si{\milli \volt}/\si{\nano\meter}), the injected current is completely refocused at the distance of 83 \si{\nano\meter} and, a periodic refocus behavior is seen in the current pattern as we further increase the fields. These oscillations in current are reminiscent of spatial Bloch oscillations or breathing modes observed in optics \cite{OpticalBloch}.
 
 We found that the revival period is inversely proportional to the field strength, cf. supplementary
  material (SM Fig.6). At high fields, confinement over the ribbons around the injection point is enhanced, and in the end the current is localizing primarily within the single injection GNR.  For the highest field here considered, we observed a faint asymmetric spill-out leakage towards the low potential region which we attribute to  Landau–Zener tunneling between the $p_z$ \textit{Wannier–Stark} ladders (cf. Fig.~\ref{fig:figure2}b). from the bulk states tunneling to the edge/pore states located at higher potentials \cite{drivenpotential}.

\section{Analytical Model of Bloch Oscillations/Breathing Modes}
To gain deeper insight into the origin of the field-induced breathing modes observed in our \gls{ls} transport simulations, we model \gls{npg} as an effective one-dimensional array of weakly coupled electron waveguides with a linear dispersion, each corresponding to an individual \gls{gnr}. 
The propagation of an injected current along the device can then be described by a discrete differential equation (DDE), formally analogous to the time-dependent Schrödinger equation in a lattice,
\begin{eqnarray}
 i\frac{d \psi_n}{dy} - \kappa( \psi_{n-1}+\psi_{n+1}) = n\Delta\,\psi_n\,.\label{eq:DDE}
\end{eqnarray}
Here, $\psi_n(y)$ denotes the amplitude of the wave-function on the $n$th \gls{gnr} at longitudinal position $y$, $\kappa$ is the inter-ribbon coupling (hopping amplitude) and $\Delta$ represents the field-induced shift in the local propagation constant between neighboring ribbons. The parameter $\alpha=\Delta/\kappa$ is a unit-less constant that quantifies the competition between field-induced localization and coupling-induced spreading. 
 
For an initial excitation localized at a single ribbon, the system exhibits the characteristic breathing dynamics. The analytical solution can be expressed in terms of Bessel functions by \cite{OpticalBloch},
\begin{eqnarray}
 \psi_n(y)=J_{-n}\!\left(\frac{4}{\alpha} \mathrm{sin}\left(\frac{\alpha}{2}y\right ) \right)\exp\!\left(i\frac{n}{2}(\alpha y- \pi )\right)\,.\label{eq:sol}
\end{eqnarray}
These solutions describe the periodic expansion and refocusing of the wavefunction along the transverse ribbon direction. This phenomenon is directly analogous to Bloch oscillations in crystal lattices under constant electric fields. Only this is typically considered in the time-domain where $y$ is replaced by time, $t$, in Eq.~\ref{eq:DDE}. 

The motion is characterized by the recurrence length, or Bloch period, given by, 
\begin{eqnarray}
  Z_T=2\pi/\Delta\,.\label{eq:ZT} 
\end{eqnarray}
At this distance, the transverse wavefunction revives its original spatial profile, giving rise to the breathing behavior seen in the transmission map. The maximum spatial extent of the wavefunction, that is, the number of GNRs involved during expansion, can be estimated as $w\approx \pm 4/\alpha$, where the transverse spread is inversely proportional to the field strength (via $\alpha$). This indicates that stronger fields confine tightly the injected current, consistent with the numerical results \cite{OpticalBloch}.

To quantitatively compare the analytical model to our  \gls{ls}  simulation, we extracted the effective parameters from the DFT calculations, as explained in the Supplementary Material. The parameters $\kappa=0.01$ 1/\AA, and using an effective band-slope (velocity) of $s^{\textbf{eff}}=3.23$ e\si{V}\AA, yielded excellent agreement for the oscillation period and the spatial distribution.
We numerically solved the DDE for a system with a total number of $n=2N_x$ sites, with $N_x=21$ corresponding to the same number of unit cells as in the calculations. \gls{ls}. We used a step size of  $\delta y=0.1$ \AA, varying the applied field from 0 to 58 \si{\milli \volt}/\si{\nano\meter}. The initial condition here employed corresponds to $\psi_n(y_0)=\delta_{n,21}$. 
The resulting intensity distributions $|\psi_n|^2$ reproduce well the key features of the \gls{ls} transport simulations, including the field-dependent periodicity and confinement. The comparison is shown in Figure \ref{fig:Bloch oscillations}.
This model not only corroborates the existence of Bloch oscillations in NPG, but also provides an intuitive framework to predict and design field-tunable transport behaviors.

For larger electrostatic doping, but still on the linear-dispersion regime of the conduction bands, Bloch period is modulated by the changes in the inter-ribbon coupling $\kappa$, as well as the local potential changes on the atoms due to charge redistribution (see Suppl. Mat.). This enabled the close comparison to the DDE model 

\section{Robustness against Structural Disorder}
To assess the experimental feasibility of observing breathing modes we finally investigate the robustness of the Bloch ocillation/breathing mode pattern in the current flow in the presence of static disorder.  As a simple model, we consider carbon vacancies. At the level of tight-binding modeling, vacancies have shown to be a good qualitative model for describing also scattering by f.ex. hydrogen adatoms on the surface of graphene \cite{hydrgoen}. However, we will focus on spin independent transport, although we note that these defects are known to give rise to local magnetic moments and thus could be interesting towards spintronic devices \cite{quantum,Han2014}.

\begin{figure}[ht]
\centering
\includegraphics[width=\linewidth]{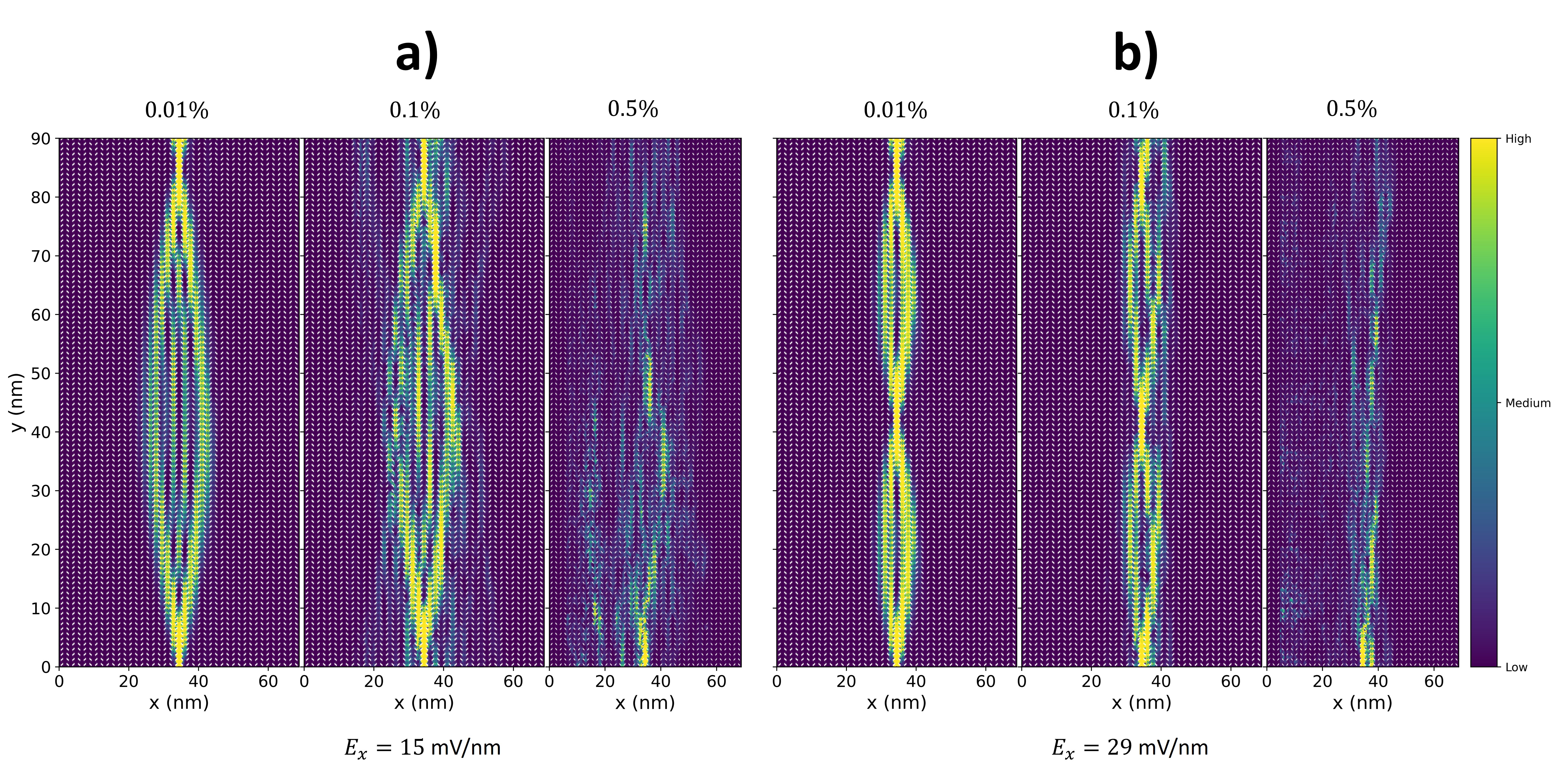}   
\caption{\textbf{Breathing Modes with disorder}. Atom transmission maps $T_i$ of \gls{npg}  with applied transverse fields of  (a) $E_x=15$ \si{\milli\volt}/\si{\nano\meter} and (b)  $E_x=29$ \si{\milli\volt}/\si{\nano\meter}. Defects, corresponding to carbon vacancies, are randomly distribute along the device region. These are shown from left to right in order of increasing concentration  from $0.01\%$ to $0.5\%$.}
\label{fig:disorder}
\end{figure}

We sampled vacancies in concentrations of $0.01 \%$, $0.1 \%$ and $0.5 \% $. Vacancy concentrations are given by the number of vacancies $N_{\text{vac}}$ relative to the number of total atoms $N$ in the pristine device, given by $(N_{\text{vac}}/N)\times
100\%$.  The removal of carbon atoms was done globally at both  bridge, and  "bulk" ribbon atoms. Concentrations of $0.1 \%$ are of interest, as have been shown to describe the charge mobilities in graphene of 2000 \si{\centi\meter}²/\si{\volt \sec}  at low temperatures (20 K) similar to experiments \cite{anisotropy}. The results are presented in Figure \ref{fig:disorder}. For samples, with the lowest vacancy concentration,  $0.01\%$,  Bloch oscillations remain intact. For a disorder of $0.1 \%$ we see a clear perturbation of the spatial features. Nevertheless, the Bloch period revival of the initially injected current is preserved. For highly disordered systems (0.5\%), the spatial patterns underpinning \gls{npg} are completely washed out by the scatterers, and the revival of the current is lost. 
These results suggest that Bloch oscillations in NPG are rather robust even for significant levels of static disorder, supporting their potential observability in experimental devices. Fine details of the wave interference behavior can be partially hampered by dynamic disorder, which is out of the scope of this work.

\section{Conclusion}
In this work, we used a \gls{ls} Non-Equilibrium Green's function transport methodology with parameters from \gls{dft} to study point-injected current flow patterns in \gls{npg} under an applied transverse electrical field. We demonstrated the emergence of Bloch oscillations/breathing modes on the 100nm-scale in \gls{npg} devices for experimentally relevant fields of few \si{\milli \volt}/\si{\nano\meter}. Bloch oscillations showed to be robust against static-disorder caused by randomly distributed vacancy-defects. The results were well explained by a simple generic waveguide model, illustrating the close analog to optics. We showed how applied fields provide a way to control the current flow in the \gls{npg} structures. Importantly, this approach relies on non-invasive electrical control. Together with controlled chemical \cite{ElectrochemicalNPG} and structural changes \cite{Bridgetune-MorenoJACS2023} in \gls{npg} this can provide a flexible route to spatially configurable graphene-based nano-circuitry. 

This work provides insights into field-induced modulation of electron transport in graphene-based architectures for applications in nano-scale electronics, quantum signal processing, and high-frequency oscillators. For example, Bloch oscillations could serve as a mechanism to encode, process, or transport quantum information within graphene-based quantum networks \cite{QuantumStateTransfer}. Specifically, GNRs could potentially act as the building block for performing controlled quantum walks \cite{Quantumwalks}, required in quantum simulation and quantum computing on solid-state 2D materials.

\section{Acknowledgments}
AAM was funded by Independent Research Fund Denmark, Grant 0.46540/3103-00229B. We acknowledge support from  the Novo Nordisk Foundation Data Science Research Infrastructure 2022 (NNF22OC0078009),  and DTU Computing Center resources, http://dx.doi.org/10.48714/DTU.HPC.0001.

\newpage
\section{Supplementary Material}
\subsection{Fitting of DDE model}
Previously  \cite{TalbotNPG}, it was found that the interribon coupling ($\kappa$) can be obtained from the band-structure of \gls{npg}:
\begin{eqnarray}
\kappa=\frac{|k_2-k_1|}{4}
\end{eqnarray}
Here, we found that the field-induced shift ($\Delta$) can be obtained  from \textit{ab-initio} \gls{dft} calculations as well,
\begin{eqnarray}
\Delta(E)=\frac{U}{2N_x}\left(\left| \frac{dE_n}{dk} \right|_{E} \right)^{-1}=e     E_x d\left(\left| \frac{dE_n}{dk} \right|_{E} \right)^{-1}\label{eq:Delta}
\end{eqnarray}
 Here, $d$ denotes the distance between adjacent GNRs. The factor of $U/2N_x$ corresponds to the net change in the on-site potential from neighboring ribbons, while the term $s_{\text{avg}}=|dE_n/dk|_E$ denotes the average slope of the conduction bands at the injection energy $E$, and corresponds to the average group velocity. For a fixed device length  of $L_x=2N_x$ and  total potential change $U$, the discrete model predicts that the current at an injection energy $E$ will  self-repeat at distances  given by the Bloch period $Z_T$:  
\begin{eqnarray}
     Z_T=2\pi\times \left( \frac{2N_x}{U}\right)\times \left| \frac{dE_n}{dk}\right|_{E}\label{eq:ZT}  
\end{eqnarray}
We found that the parametrization of $\Delta$ with the average band-slope ($s_\text{avg}$) provides a good estimate for the Bloch period $Z_T$ observed in the large scale calculations.   

\begin{figure}[ht]
\centering
\includegraphics[width=\linewidth]{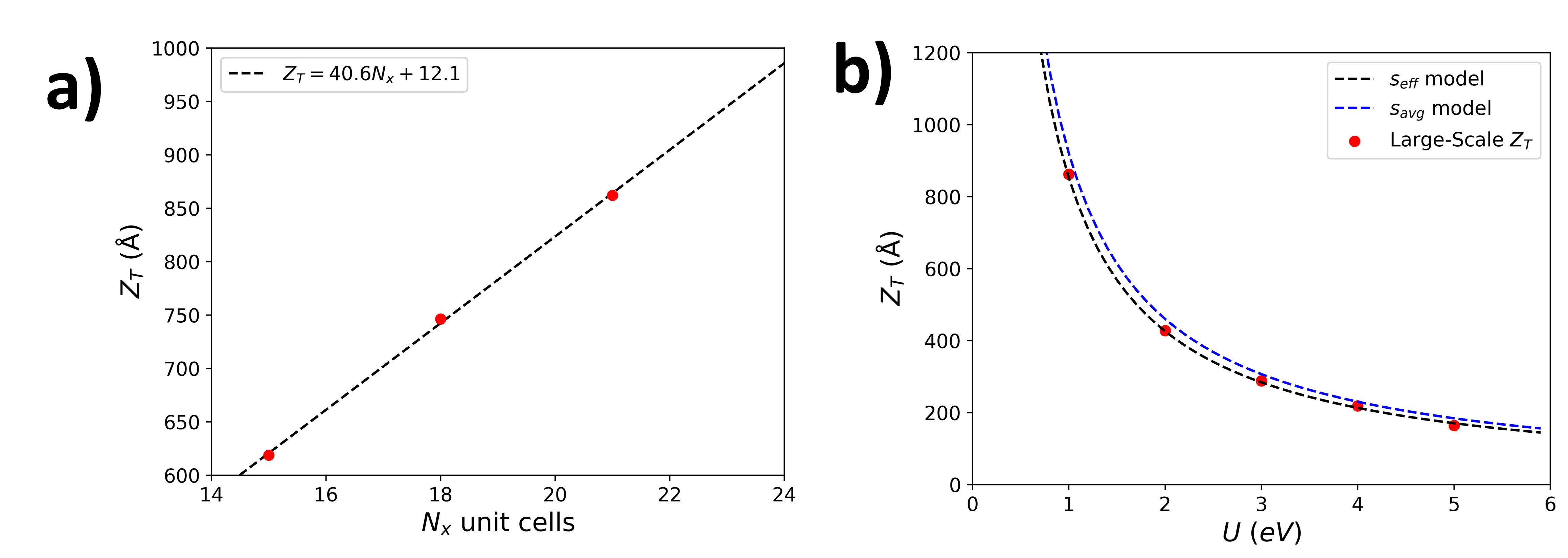} 
\caption{\textbf{Existence of Breathing Modes on the linear dispersion regime.}(a) Linear relation between the Bloch period ($Z_T$) and the number of unit cells along x direction ($N_x$). The potential drop   is fixed at $U=1$ e\si{\volt} along $x$ in the device. b) $Z_T$ exhibiting $1/U$ dependency on \gls{ls} calculations. In red dots, the observed $Z_T$ obtained with the \gls{ls} method are show. The dashed black line corresponds to the prediction  from the $s^{\text{eff}}$ parametrization. The dashed blue line, corresponds to the prediction from the average band-slope  $s^{\text{avg}}$.}\label{fig:Linear}
\end{figure}

We tested the linear dependency of $Z_T$ with the device length by  performing a series of \gls{ls} calculations, fixing $U=1$  e\si{ \volt}, and increasing the device length $N_x$. This allowed us to parametrized an effective band-slope  $s^{\text{eff}}=3.23$ e\si{\volt}\AA, which we can compare to graphene $s^{\text{Gr}}\sim 5.5$ e\si{\volt}\AA. This is shown in Figure \ref{fig:Linear}a. We used this effective band-slope to calculate $\Delta$ and obtained a parameterized DDE that describes the $Z_T$ observed in the Bloch oscillations (Fig.\ref{fig:Linear}b). We observed excellent agreement between the DDE parameterized with $s_{\text{eff}}$ (dashed black line)  and the LS results (red dots). Slightly higher periods are observed for parameters fitted  from \gls{dft} band-slope instead, setting $s^{\text{av}}=3.49$ e\si{\volt}\AA  (dashed blue line).

\subsection{Gate dependence on Bloch oscillations}
\begin{figure}[ht]
\centering
\includegraphics[width=\linewidth]{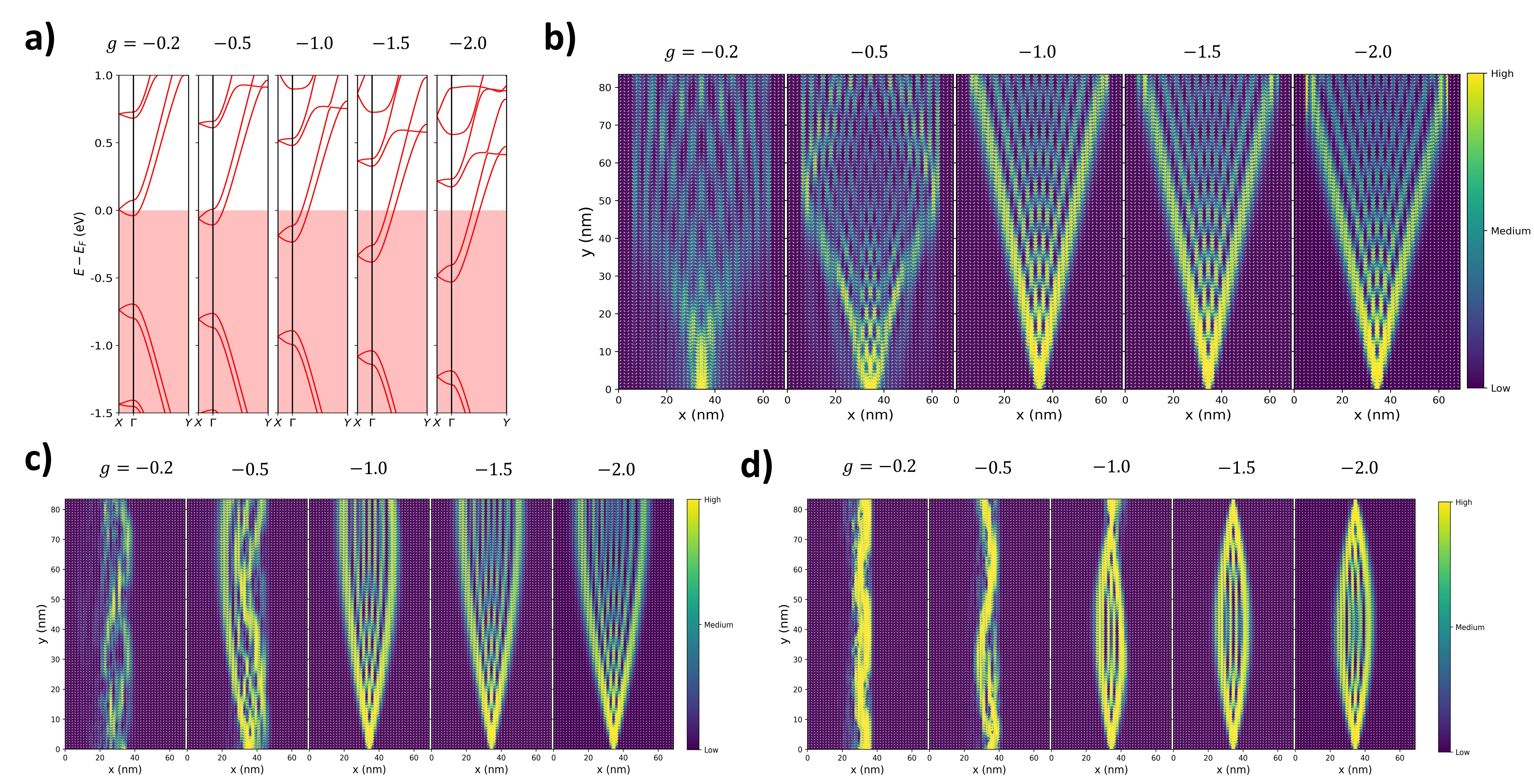}
    \caption{\textbf{Dependency of Bloch Oscillation on electrostatic gating} a) Band structure from parameter-free TB Hamiltonian of \gls{npg}  under different electrostatic gating ($g=-0.2,-0.5,-1.0,-1.5,-2.0$). Atom transmission maps $T_i$ for \gls{npg} devices with fields of  $E_x=0$ (b), $7$ (c) and $15$ (d)  \si{\milli\volt}/\si{\nano\meter} with increasing electrostatic gating. Injection  correspond to $E=E_F$.}
    \label{figa:gate_bo}
\end{figure}

We investigated the dependency of breathing modes from Bloch oscillations on the applied gate. We considered \gls{tb} $p_z$-pruned Hamiltonians from DFT calculations on $n$-doped \gls{npg}s with different gating concentrations ( $g=-0.2,-0.5,-1.0,-1.5,-2.0$ ). The electronic bands are shown in Figure \ref{figa:gate_bo}a. The first case corresponds to a doping where only the lowest conduction band is filled. For the second case ($g=-0.5$), both bands are filled, but the \gls{ef} lies at energies of low dispersion ($\Gamma\rightarrow X$) in the second conduction band. For the rest of the cases ( $g=-1.0,-1.5,-2.0$ ), both bands are filled, and the bands exhibit linear dispersion along $\Gamma\rightarrow Y$.

We constructed \gls{ls} transport devices with a field of $E_x=0.7$ \si{\milli \volt}/\si{\nano\meter} as described in  Section 2.2.  Single-point injection at the Fermi energy ($E=E_F$) was considered. We plotted the atom transmission maps in Figure \ref{figa:gate_bo}. For a gate of $g=-0.2$ only the lowest band is available at the $E_F$ and is approximately parabolic. Complex scattering takes place and the Bloch oscillations are not present for non-zero fields. At $g=-0.5$ both bands present propagating states at $E_F$ around the injecting ribbon. Here some signatures of Bloch oscillation appear, but with a distortion due to the field-induced energy up-shift with increasing $x$, which render the local ribbon bands non-conducting, and break left-right-symmetry.  Only when both bands are filled and stay in the linear band regime along $\Gamma\rightarrow Y$, Bloch oscillations become clear. 

An increase in $Z_T$ and the width of the oscillation amplitude, $w$, is observed with increasing gate. The change of $Z_T$  occurs due to (i) the local changes in the on-site potentials at the carbon atoms as the electrostatic gate increases, and (ii) the change in the group velocity of the bands.

\end{document}